\documentclass[sigconf,anonymous=false]{acmart}
\AtBeginDocument{%
  }

\setcopyright{acmlicensed}
\copyrightyear{2025} 
\acmYear{2025} 
\setcopyright{cc}
\setcctype{by}
\acmConference[RecSys '25]{Proceedings of the Nineteenth ACM Conference on Recommender Systems}{September 22--26, 2025}{Prague, Czech Republic}
\acmBooktitle{Proceedings of the Nineteenth ACM Conference on Recommender Systems (RecSys '25), September 22--26, 2025, Prague, Czech Republic}\acmDOI{10.1145/3705328.3748148}

\acmISBN{}

\usepackage{color,soul}
\usepackage{colortbl}

\newcommand\width{0.8}
\newcommand\framework{Informfully Recommenders}
\newcommand\drdw{D-RDW}
\newcommand\rdw{$RP^{3}_{\smash{\beta}}$}

\begin{document}

\title[Informfully Recommenders]{Informfully Recommenders -- Reproducibility Framework\\for Diversity-aware Intra-session Recommendations}

\author{Lucien Heitz}
\email{heitz@ifi.uzh.ch}
\orcid{0000-0001-7987-8446}
\affiliation{%
  \institution{Department of Informatics,\\University of Zurich}
  \streetaddress{Binzmühlestrasse 14}
  \city{Zurich}
  \country{Switzerland}
  \postcode{CH-8050}
}

\author{Runze Li}
\email{runze.li@uzh.ch}
\orcid{0009-0001-7284-753X}
\affiliation{%
  \institution{Department of Informatics,\\University of Zurich}
  \streetaddress{Binzmühlestrasse 14}
  \city{Zurich}
  \country{Switzerland}
  \postcode{CH-8050}
}

\author{Oana Inel}
\email{inel@ifi.uzh.ch}
\orcid{0000-0003-4691-6586}
\affiliation{%
  \institution{Department of Informatics,\\University of Zurich}
  \streetaddress{Binzmühlestrasse 14}
  \city{Zurich}
  \country{Switzerland}
  \postcode{CH-8050}
}

\author{Abraham Bernstein}
\email{bernstein@ifi.uzh.ch}
\orcid{0000-0002-0128-4602}
\affiliation{%
  \institution{Department of Informatics,\\University of Zurich}
  \streetaddress{Binzmühlestrasse 14}
  \city{Zurich}
  \country{Switzerland}
  \postcode{CH-8050}
}

\renewcommand{\shortauthors}{Heitz et al.}

\begin{abstract}
Norm-aware recommender systems have gained increased attention, especially for diversity optimization.
The recommender systems community has well-established experimentation pipelines that support reproducible evaluations by facilitating models’ benchmarking and comparisons against state-of-the-art methods.
However, to the best of our knowledge, there is currently no reproducibility framework to support thorough norm-driven experimentation at the pre-processing, in-processing, post-processing, and evaluation stages of the recommender pipeline.
To address this gap, we present \framework{}, a first step towards a normative reproducibility framework that focuses on diversity-aware design built on Cornac.
Our extension provides an end-to-end solution for implementing and experimenting with normative and general-purpose diverse recommender systems that cover
1) dataset pre-processing,
2) diversity-optimized models,
3) dedicated intra-session item re-ranking, and
4) an extensive set of diversity metrics.
We demonstrate the capabilities of our extension through an extensive offline experiment in the news domain.
\end{abstract}


\maketitle

\section{Introduction}
\label{sec:introduction}
Recommender systems (RSs) help users to find their way in the vast online information space, shape the public discussion, and serve as a foundation for public cohesion~\cite{helberger2018exposure,bernstein2021diversity,heitz2022benefits}.
In the news domain, for example, they fulfill an important ``democratic role''~\cite{helberger2019democratic} for political opinion formation and 
informational self-determination~\cite{sargeant2022spotlight}.
\newpage

Therefore, depending on the target domain, RSs face unique challenges that require balancing societal norms and values on the one hand~\cite{helberger2018exposure} as well as technical performance and economic goals of platform owners on the other hand~\cite{cohen2002online,richardson2007predicting}.
In the context of this paper, we refer to such RS instances as a \textit{normative} RS (NRS).
We follow the definition of~\citet{vrijenhoek2023normalize} where normativity is understood as the practice of operationalizing societal norms and values as part of the RS pipeline. 

Development and evaluation of NRSs is challenging.
When looking at a target objective for NRSs, such as diversity, there is disagreement on the conceptual level on the precise notion~\cite{loecherbach2020unified,vrijenhoek2024diversity}.
Additionally, online user studies for assessing the algorithm's impact on users remain the exception in RS research~\cite{bauer2024values}, and the understanding of NRSs as well as re-rankers remains limited~\cite{treuillier2022being}.
These studies, however, are vital to assess the performance of NRSs, because with the predominant focus on offline evaluation, it is unclear how algorithms impact users~\cite{jannach2020escaping}.

One reason for this lack of proper assessment is the requirement to have sufficiently rich datasets with complementary information on items, participants' backgrounds~\cite{iana2023nemig,heitz2024idea}, normative models/re-rankers~\cite{castells2021novelty}, diversity metrics~\cite{vrijenhoek2024diversity}, and visualizations~\cite{beel2021unreasonable,heitz2024informfully} for the evaluation with users.
Despite recent efforts to promote more normative and beyond accuracy perspectives (e.g., see RecSys 2024 Challenge~\cite{kruse2024recsys,kruse2024eb,heitz2024recommendations}), there have not yet been any dedicated end-to-end pipelines proposed for the systematic evaluation of NRS.
To address these shortcomings, we present \framework{}, a first open-source reproducibility framework for norm-aware approaches, such as diversity.\footnote{Informfully Recommenders: \url{https://github.com/Informfully/Recommenders}\\ Experiment configuration files: \url{https://github.com/Informfully/Experiments}\\ We provide a complementary online documentation of the codebase with an extended technical description of each of the pipeline stages that we outline in this paper: \url{https://informfully.readthedocs.io}} 

Our framework's contributions include:
1) six out-of-the-box dataset augmentation functions to add norm-relevant dimensions to items (supporting multiple languages),
2) three random walks and two lightweight diversity models for norm-aware recommendations,
3) three diversity-optimized re-ranker algorithms for use with existing models, together with two intra-session re-rankers for the user simulator,
4) five traditional and five normative diversity metrics for assessing recommendations, and
5) compatibility with the Informfully Research Platform ~\cite{heitz2024informfully} to support item visualization for online user studies.
It is designed as an extension to the well-established Cornac framework~\cite{salah2020cornac,truong2021multi}, providing an end-to-end solution for implementing and experimenting with NRSs.

\begin{table*}

    \caption{Overview of open-source reproducibility framework. The comparison looks at supported datasets, models, re-rankers, and metrics, as well as augmentation, simulation, and visualization capabilities.}
    \scalebox{0.945}{
       
        \begin{tabular}{l|c|ccc|ccc}

            \toprule
            
            Framework &
                \parbox{1.73cm}{\centering{Modes}} &
                \parbox{1.73cm}{\centering{Models}} &
                \parbox{1.73cm}{\centering{Re-rankers}} &
                \parbox{1.73cm}{\centering{Metrics}} &
                \parbox{1.7cm}{\centering{Data Aug-mentation}} &
                \parbox{1.7cm}{\centering{User Simulator}} &
                \parbox{1.7cm}{\centering{Item Visualization}} \\
            
            \midrule

            ClayRS~\cite{lops2023clayrs} &
                OFF &
                TRA &
                 &
                ACC &
                &
                &
                \\
  
            Cornac~\cite{salah2020cornac,truong2021exploring,truong2021multi} + A/B~\cite{ong2024cornac} &
                ON, OFF &
                TRA &
                 &
                ACC, BEY &
                &
                &
                $\checkmark$ \\
        
            daisyRec 2.0~\cite{sun2022daisyrec} &
                OFF &
                TRA &
                 &
                ACC, BEY &
                &
                &
                \\
        
            Elliot~\cite{anelli2021elliot} &
                OFF &
                TRA &
                 &
                ACC, BEY &
                &
                &
                \\
        
            FuxiCTR~\cite{zhu2021open,zhu2022bars} &
                OFF &
                TRA &
                 &
                ACC &
                &
                &
                \\

            LensKit~\cite{ekstrand2020lenskit} &
                OFF &
                TRA &
                 &
                ACC &
                &
                &
                \\
                
            Microsoft Recommenders~\cite{argyriou2020microsoft} &
                OFF &
                TRA &
                 &
                ACC, BEY &
                &
                &
                \\

            RecBole~\cite{zhao2021recbole} &
                OFF &
                TRA &
                 &
                ACC, BEY &
                $\checkmark$ &
                &
                \\
        
            ReChorus 2.0~\cite{li2024rechorus2} &
                OFF &
                TRA &
                STA &
                ACC &
                &
                &
                \\
        
            RecList~\cite{chia2022beyond} &
                OFF &
                TRA &
                 &
                 &
                &
                $\checkmark$ &
                \\
        
            RecPack~\cite{michiels2022recpack} &
                OFF &
                TRA &
                STA &
                ACC, BEY &
                $\checkmark$ &
                &
                \\

            \midrule

            \framework{} &
                ON, OFF &
                NOR, TRA &
                STA, DYN &
                ACC, BEY &
                $\checkmark$ &
                $\checkmark$ &
                $\checkmark$ \\
        
            \bottomrule
            
        \end{tabular}

    }

    \Description{Overview of open-source reproducibility framework. The comparison looks at supported datasets, models, re-rankers, and metrics, as well as augmentation, simulation, and visualization capabilities.}
    \label{tab:comparison}
    
\end{table*}

This enables \framework{} to assist the systematic integration, evaluation, and assessment of societal values and norm-awareness into recommender systems.
We show the applicability of our extension in the context of news recommendations---a domain closely linked to normative societal values in general~\cite{helberger2019democratic,vrijenhoek2021recommenders,vrijenhoek2023normalize} and diversity in particular~\cite{vrijenhoek2022radio,heitz2024recommendations}.


\section{Related Work}
\label{sec:related_work}
In this section, we compare RS reproducibility frameworks to show their respective shortcomings for assessing NRSs.
Table~\ref{tab:comparison} presents a comparison of open-source reproducibility frameworks that can be used to tackle news recommendations and diversity.\footnote{This list is based on the ACM RecSys repository of evaluation framework recommendations: \url{https://github.com/ACMRecSys/recsys-evaluation-frameworks}}
We compare the capabilities of the frameworks on the following dimensions:
\begin{description}
    \item [Modes:] 
    Shows if frameworks support \textbf{on}line user experiment (ON) or if they are focusing on \textbf{off}line benchmarking (OFF).
    \item [Models:]
    Lists available model types.
    Options include \textbf{nor}mative (NOR) and \textbf{tra}ditional models (e.g, accuracy, TRA).
    \item [Re-rankers:]
    We differentiate between \textbf{sta}tic re-ranking of candidate lists after the model stage (STA) and \textbf{dyn}amic intra-session re-ranking that takes user interactions into account and can iterate multiple times (DYN). The re-ranker must be part of a modularized pipeline. Re-rankers tied to/part of a model cannot be reused across different algorithms. In both cases we leave the entry empty.
    \item [Metrics:]
    Shows if generic \textbf{acc}uracy (ACC) or norm-relevant \textbf{bey}ond accuracy metrics (BEY) are included.
    \item [Augmentation, Visualization, Simulator:]
    Furthermore, our comparison covers three additional dimensions critical for NRSs:
    1) data augmentation (to add required normative attributes for models, re-rankers, or metrics), 
    2) user simulator (for offline benchmarking of intra-session re-ranking), and 
    3) item visualization (for conducting user studies).
    We mark an entry with $\checkmark$ if it includes data augmentation, a user simulator, or item visualization.
\end{description}

\begin{figure*}[!ht]
    \centering
    \includegraphics[width = 170mm]{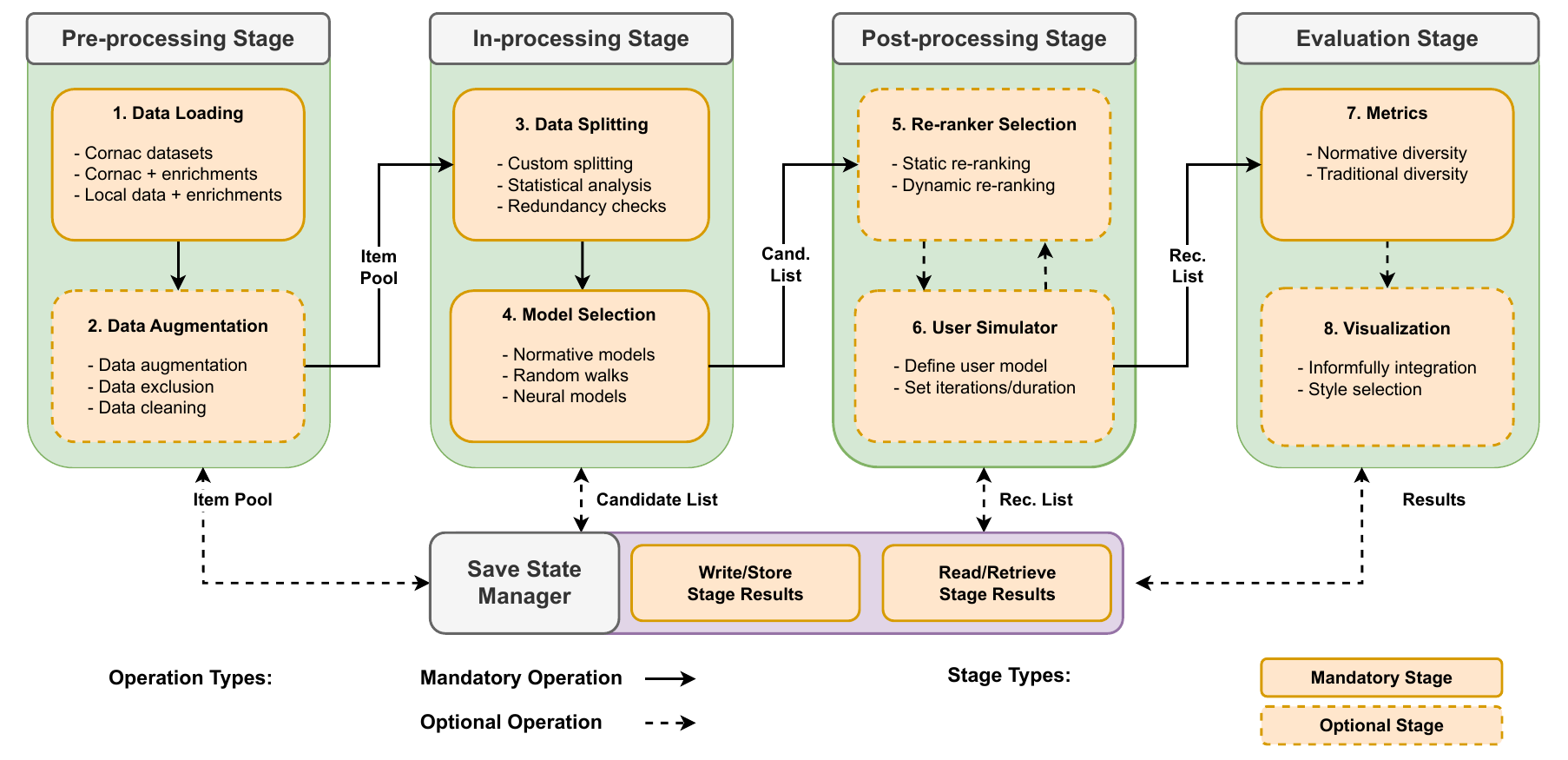}
    \caption{\framework{} extension of the existing Cornac pipeline by implementing a diversity-aware four-stage RS pipeline with eight customizable steps. It includes a Save State Manager for saving and loading results at each stage. Information gets passed across stages with specific files (i.e., item pool, candidate lists, and recommendation lists).}
    \label{fig:overview}
    \Description{TBD}
\end{figure*}

Looking at Table~\ref{tab:comparison}, we see that the support of online experimentation is limited to one framework.
We do not find any support for norm-aware models.
The same goes for re-ranking, where all but one framework do not even include a dedicated stage for this process.
The situation is better for metrics, where six frameworks include beyond accuracy assessment that consider diversity.

\paragraph{\textbf{Data Augmentation:}}
Experimentation with NRSs is heavily dependent on rich contextual information, as models, re-rankers, or metrics require, e.g., information on political actors or item embeddings to work.
This data, however, is rarely present in datasets.
Table~\ref{tab:comparison} shows that only two frameworks offer built-in functionality for data augmentation to add such norm-relevant information.

\paragraph{\textbf{User Simulators:}}
Intra-session effects can have a significant impact on user engagement~\cite{lu2019effects}.
Using sequence-aware information on intra-session behavior offers a rich source of information to personalize recommendations~\cite{symeonidis2022sequence}.
However, leveraging intra-session data for news recommendations is among the least popular domains in sequence-aware RS research~\cite{quadrana2018sequence}.

While there exists previous work on user simulation for recommender system, they are either conceptual or theoretical in nature without implementation~\cite{hazrati2022simulating}, stand-alone implementations that are not part of any testing framework~\cite{ie2019recsim,zhao2023kuaisim,chen2023sim2rec}, tied to a specific domain~\cite{afzali2023usersimcrs}, or a combination of simulators and re-rankers~\cite{yao2021measuring}.
We see this also reflected in Table~\ref{tab:comparison}, where only a single framework offers the capability of testing with simulated user interactions.

\paragraph{\textbf{Item Visualization:}}
Only the extended Cornac A/B supports item visualization for user experiments.
While we use the same underlying Cornac back end, our extension utilizes Informfully~\cite{heitz2024informfully}, a more generic approach that is not tied to a specific framework.
This allows our framework to act as a back end for supporting online user studies by taking the recommendation lists and forwarding them to an application where people can interact with them.

To the best of our knowledge, there is no open-source reproducibility framework that offers a unified, end-to-end solution across the four main RS stages with dataset operations, model selection, (intra-session) re-ranking with user simulation, and metrics assessment together with item visualization.

\section{\framework}
\label{sec:framework}
\framework{} is an extension of the Cornac framework for multimodal RS~\cite{salah2020cornac,truong2021exploring,truong2021multi}.
Figure~\ref{fig:overview} provides an overview of the updated reproducibility framework.
Following the modularized stages of Table~\ref{tab:comparison}, our norm-aware diversity extension of Cornac splits the RS pipeline into the four stages of
pre-processing (dataset operations),
in-processing (model operations),
post-processing (re-ranking operations),
and evaluation (metrics, assessment, and visualization).
Each stage is further subdivided into two steps, allowing researchers to customize the intended behavior of the RS.

Communication between stages is done exclusively via exchanging item files (i.e., Item Pool, Candidate List, and Recommendation List---solid arrows in Figure).
We also provide a Save State Manager for storing and retrieving these item files.
This manager allows the pipeline to be initialized at any stage by reusing existing intermediate results, speeding up the development process (e.g., when testing multiple re-ranking approaches, one candidate list is sufficient as it can be (re-)loaded for subsequent re-ranking rounds, skipping the pre- and in-processing stages).

Our extended framework presents a complete end-to-end pipeline for RS.
It allows for reproducibility across the in-processing~\cite{wan2023processing}, post-processing~\cite{petersen2021post}, and the evaluation stage~\cite{heitz2023classification}.
These capabilities allow \framework{} to be used for both general-purpose and diversity-driven offline benchmarking/development purposes, as well as being deployed as a back end for conducting user studies.
It has a successful track record of powering online user studies (for more details, please see~\cite{heitz2022benefits,heitz2023deliberative,heitz2024recommendations,heitz2024idea}).
Furthermore, by making the newly introduced stages and steps optional, we ensure full backwards compatibility with existing Cornac experiments. 

\subsection{Pre-processing Stage}
\label{sec:stage-1}
The purpose of the pre-processing stage is to prepare the user-item interactions and to define the item pool.
The extension to the pre-processing stage includes two main additions:
1) customizable data loading options and
2) data augmentation functions.

\paragraph{\textbf{Data Loading:}}
Cornac uses a user-item rating matrix 
to load and process data, which 
does not contain any information on when an interaction took place.\footnote{This time component, however, can be crucial, as it allows, e.g., discounting older interactions and/or tracking the status of impression lists.}
We now allow users to load separate history files during the in-processing stage via the $userHistory$ parameter that can contain custom attributes.

Looking at the recommendation output, Cornac considers items that appear both in the training \textit{and} the test sets, or items from the test set alone for recommendation purposes.
However, this can be too broad or too narrow depending on the specific use case.
We, therefore, extended the base recommendation model 
and re-ranker 
with an optional $articlePool$ parameter to either extend or reduce the item pool for which predictions are calculated.
The parameters $userHistory$ and $articlePool$ are optional, making the extension fully backwards compatible with existing Cornac experiments.

\paragraph{\textbf{Data Augmentation:}}
The data augmentation extension is an optional step comprising several \textit{text} enrichment functions, such as sentiment analysis, named entity recognition (NER), the identification of political actors and parties, assessment of text complexity, identification of event clusters, and categorization of item types. 
The data augmentation pipeline supports texts in multiple languages, including English, German, Danish, and Portuguese.

\begin{description}
    \item [Sentiment Analysis:]
    We offer sentiment analysis for texts using RoBERTa.\footnote{The underlying model can be exchanged. Our sample implementation uses XLM-roBERTa: \url{https://huggingface.co/cardiffnlp/twitter-xlm-roberta-base-sentiment}}
    The sentiment ranges from negative ($-1.0$) to positive ($1.0$).
    We take this score and group articles into four baskets, expressing an opinion that is either ``negative,'' ``somewhat negative,'' ``somewhat positive,'' or ``positive.''\footnote{The number of baskets used here only serves as an example and can be customized.}
    \item [Named Entity Recognition:]
    Named entities of various types (e.g., people, locations, organizations, events, among others) are extracted using the spaCy library.\footnote{NER with spaCy: \url{https://spacy.io/usage/linguistic-features\#named-entities}}
    \item [Political Actors:]
    The political augmentation identifies politicians and parties using a combination of spaCy for NER and Wikidata\footnote{Wikidata website: \url{https://www.wikidata.org}} for further augmenting the named entities with politics-centric information. 
    The script detects party names in the text.
    Afterwards, labels for ``Governing Party,'' ``Opposition Party,'' or ``Others/Foreign Parties'' can be assigned using a custom mapping provided by the user.
    \item [Text Complexity:]
    The framework assesses the complexity of a text using the Textstat library.\footnote{Textstat library: \url{https://pypi.org/project/textstat}}
    \item [Story Clusters:]
    We calculate story clusters to allow grouping texts, such as news articles, by events based on text similarity and named entities within categories using NetworkX.\footnote{NetworkX documentation: \url{https://python-louvain.readthedocs.io}}
    \item [Article Categories:]
    This augmentation feature allows for automatically assigning a category to a text using BART.\footnote{BART model: \url{https://huggingface.co/facebook/bart-large-mnli}}
    \item [Helper Function:]
    We include auxiliary functions to clean and validate the original as well as the augmented data files.
    Options include filtering invalid articles (i.e., items with empty attributes) and removing users or items with no recorded interactions/history.
    After validating all the articles, the script can prepare the user-item rating matrix required by Cornac.
\end{description}

We provide ready-to-go augmentation pipelines and sample code for the Ekstra Bladet News Recommendation Dataset (EB-NeRD, Danish)~\cite{kruse2024eb}, the Microsoft News Dataset (MIND, English)~\cite{wu2020mind}, and the German News Collection on Migration (NeMig, German)~\cite{iana2023nemig} to showcase how the added augmentation steps perform across different languages and datasets.\footnote{The code is available: \url{https://github.com/Informfully/Experiments}}

\subsection{In-processing Stage}
\label{sec:stage-2}
The purpose of the in-processing stage is to generate a candidate list of items from the item pool.
The extended framework contains five new data splitting methods and three new families of algorithms that allow for experimenting with both diversity-driven recommendations and news recommendations.

\paragraph{\textbf{Data Splitting:}}
We introduce five additional data splitting methods:
attribute-based sorting,
diversity-based subset construction,
attribute-based stratified splitting,
diversity-based stratified splitting, and
a clustering-based approach.
The main motivation behind these splitting methods is not primarily the improvement of target metrics.
Instead, the goal is to see how the model's performance is affected by changes in the underlying dataset.

\begin{description}
    \item [Attribute-based Sorting:]
    Allows sorting by item or user attributes before splitting, e.g., by article sentiment, to see how the resulting recommendation changes if the training set mainly consists of articles of a particular sentiment.
    \item [Diversity-based Subset Construction:]
    Construction of an item subset for training and testing with a purposefully skewed diversity across a target dimension to ascertain how this imbalance impacts recommendations. 
    \item [Attribute-based Stratified Splitting:]
    Stratified splitting that allows for the generation of train and test sets with balanced item attributes (e.g., equal distribution of political parties).
    \item [Diversity-based Stratified Splitting:]
    Measures users' diversity (e.g., in terms of political viewpoints) and controls their distribution across training and test sets.
    \item [Clustering-based Stratified Splitting:]
    Using K-means and PCA clustering approaches to control the homogeneity of training and test sets.
\end{description}

\paragraph{\textbf{Model Selection:}}
Our framework extends Cornac with three families of algorithms:
1) five neural models from the literature, 
2) our norm-aware filtering algorithms (for both on- and offline use), and 
3) three random-walk-based approaches.

\begin{description}

    \item[Neural Models:]
    Our extension includes five \textbf{non-normative} neural baseline models from past RS challenge tasks~\cite{wu2020mind,kruse2024eb}.
    We included the Efficient Neural Matrix Factorization (ENMF)~\cite{chen2020efficient}, Long- and Short-Term User Representation (LSTUR)~\cite{an2019neural}, Neural News Recommendation with Personalized Attention (NPA)~\cite{wu2019npa} and with Multi-Head Self-Attention (NRMS)~\cite{wu2019nrms}, as well as variational autoencoders (VAE)~\cite{liang2018variational}.
    
    \item[Filtering Algorithms:]
    The filtering algorithms present algorithms that incorporate social norms and values by using a normative target distribution (NTD).
    An NTD is a list of item attributes, relevant attribute values, and the overall occurrence count of these values.\footnote{In the case of news, for example, the editors can define an NTD for political parties mentions (item attribute) that gives party A and B (attribute values) the same exposure ($50$ mentions of party A and $50$ mentions of party B) to ensure balanced reporting reflecting existing journalistic principles or regulatory requirements.}
    We include two lightweight, \textbf{normative} filtering algorithms from the literature, namely participative Political Diversity (PLD)~\cite{heitz2022benefits} and deliberative Exposure Diversity (EPD)~\cite{heitz2023deliberative} as outlined by~\citet{helberger2019democratic}.
    PLD implements a model of the \textit{participatory} understanding of democracy, focusing on \textit{topic diversity} by creating a set of articles on key issues shared across all users.
    %
    EPD implements a model of the \textit{deliberative} understanding of democracy, focusing on \textit{viewpoint diversity} by providing equal exposure (e.g., to different political parties).
    Both algorithms were adapted to also work in an offline setting.
    
    \item[Random Walks:]
    \label{sec:random_walks}
    Our extension includes the \textbf{non-normative} random walk algorithm with popularity discount \rdw~\cite{christoffel2015blockbusters} and Random Walks with Erasure (RWE-D)~\cite{paudel2016updatable}, as they have shown to have excellent performance in the item diversification problem.
    Finally, we include Diversity-Driven Random Walk (D-RDW)~\cite{li2025drdw}, a novel \textit{normative} algorithm that capitalizes on the diversification capabilities of the traditional random walk algorithms and combines it with NTD.
    
\end{description}

\subsection{Post-processing Stage}
\label{sec:stage-3}
The goal of the post-processing stage is to offer a lightweight re-ranking option for recommendations to accommodate metric optimization and/or business logic. 
This is a new step in the Cornac pipeline.
To make everything backward-compatible, we set default parameters for the new stages to run old experiments that do not explicitly specify a re-ranking step.
In addition to one-time re-rankers (i.e., \textit{static} re-ranking), this stage also includes a user simulator to support iterative re-ranking (i.e., dynamic re-ranking).

\paragraph{\textbf{Re-ranker:}}
\framework{} presents a fully customizable approach that allows for static (one-time heuristics/filters) and dynamic re-ranking (accounting for intra-session user interactions).
The re-ranking logic allows for intra-session adjustments of the recommendation list in a static and dynamic fashion.

\begin{description}
    \item [Static Re-rankers:]
    The static re-rankers are applied to the candidate list right after the model step to optimize the output for a target metric such as diversity~\cite{castells2021novelty}.
    To that end, the output of the models can be re-ranked using three customized approaches:
    1) Greedy-KL (G-KL)~\cite{steck2018calibrated},
    2) PM-2~\cite{dang2012diversity}, and
    3) MMR~\cite{carbonell1998use}.\footnote{G-KL and PM-2 use the same NTD as \drdw. MMR uses a combined feature vector of a one-hot encoded sentiment vector (based on predefined sentiment bins) and a one-hot encoded political party vector.}
    \item [Dynamic Re-rankers:]
    Alternatively, we implemented a dynamic intra-session re-ranking option that updates recommendations based on user interaction (using items from the candidate list).
    The default strategy implemented in this framework is the dynamic attribute penalization (DAP).
    DAP diversifies the recommendation list by penalizing items in upcoming sessions that share attributes with clicked ones.\footnote{For example, implemented a default rule that removes any items that were already clicked earlier during the session.}
\end{description}

\paragraph{\textbf{User Simulator:}}
Dynamic re-ranking requires an underlying user model that specifies how the item feed is being browsed.
We provide a sample template that can be customized and extended.
In the context of NRSs, the two default behaviors included in the framework are:
1) Users are more likely to click on articles from a category that they have previously read, and
2) Items higher up in the recommendation list are more likely to be clicked (cf.~\cite{yao2021measuring}).
Apart from the interactions, the user models allow researchers to specify the overall duration and number of loops (i.e., how many times recommendations are calculated and ``consumed'' by the agent).

\subsection{Evaluation Stage}
\label{sec:stage-4}
The evaluation stage includes the final steps of the recommendation pipeline.
The two main contributions of our extension are:
1) beyond accuracy metrics to assess the recommendation quality in terms of item diversity and 
2) item visualization to show the system output to users for conducting online experiments to gather feedback.

\paragraph{\textbf{Metrics:}}
As part of our framework, we implemented traditional diversity metrics such as intra-list distance and expected intra-list distance~\cite{bradley2001improvingRD}, Gini coefficient~\cite{castells2021novelty}, $\alpha$-nDCG~\cite{clarke2008novelty}, and binomial diversity~\cite{vargas2014coverage}.
We feature these metrics as they are among the most prominent diversity measurements from the literature, domain-agnostic, and applicable to a wide range of different use cases~\cite{castells2021novelty}.

Furthermore, we integrate five rank-aware divergence metrics for measuring \textit{normative diversity}, called the RADio metrics~\cite{vrijenhoek2022radio}, namely calibration, fragmentation, activation, representation, and alternative voices.
The normative RADio metrics are based on democracy theory~\cite{helberger2019democratic} and are tailored for assessing the normative dimension of RSs~\cite{vrijenhoek2022radio}.
They consider various item features, such as topics, sentiment, named entities, and political parties, as well as additional contextual information, such as the user history, the pool of available items, and relevance.
We included the RADio metrics in our extension as they present a first operationalization of normative aspects for measuring recommendations.
This allows us to compare and contrast normative diversity with traditional measures for a detailed assessment of trade-offs.


\paragraph{\textbf{Visualization:}}
The last step of the NRS pipeline consists of visualizing the recommendation lists, e.g., for conducting online user studies.
To that end, our framework has built-in support for the Informfully Research Platform~\cite{heitz2024informfully}.
We provide a script and tutorial\footnote{Online resources:\url{https://informfully.readthedocs.io/en/latest/recommendations.html}} to transform the item recommendations of our framework extension into Informfully's JSON recommender exchange format (JREX) to feature item recommendations in a mobile or web app.
The script includes the selection of different visualization styles to determine how items are displayed on screen.


\section{Experiments}
\label{sec:experiment}
We demonstrate the capabilities of \framework{} by running diversity-focused experiments using neural models, filtering algorithms, and random walks on reference news datasets.

\paragraph{\textbf{Datasets:}}
In our experiments, we used three well-known news datasets, namely EB-NeRD (small version)~\cite{kruse2024eb}, MIND (small version)~\cite{wu2020mind}, and NeMig (German subset)~\cite{iana2023nemig} to compare and evaluate a diverse range of models across normative and traditional diversity metrics.
Table~\ref{tab:datasets} provides an overview of the datasets. 

We limit the data cleaning steps to removing users from the test set that are not part of the training set and removing items that have empty/no text attributes (e.g., movie trailers).
We applied data augmentation steps, as outlined in Section~\ref{sec:stage-1}.
We performed NER to identify political actors and parties in articles and assign them to one of three buckets: governing party, opposition party, or others (independent and foreign parties). 
Furthermore, we perform sentiment analysis to classify each article and apply event clustering to identify articles that cover similar stories.
This is all done using the newly added data augmentation functions.

\begin{table}[!h]
    \centering
    \caption{Comparison of EB-NeRD, MIND, and NeMig in terms of users for the train and test set, the average number of articles in a user history, as well as the total number of impressions, articles, and unique article categories.}
    \scalebox{0.9}{
    \begin{tabular}{lccccccc}

        \toprule
        \parbox{1.0cm}{\textbf{Dataset}} &
            \parbox{0.9cm}{\centering{\textbf{Train Users}}} &
            \parbox{0.9cm}{\centering{\textbf{Test Users}}} &
            \parbox{0.9cm}{\centering{\textbf{History (AVG)}}} &
            \parbox{0.9cm}{\centering{\textbf{Imp.}}} &
            \parbox{0.9cm}{\centering{\textbf{Art.}}} &
            \parbox{0.9cm}{\centering{\textbf{Cat.}}} \\
        \midrule
        MIND &
            49,823 &
            48,592 & 
            21.68 & 
            7,336,094 & 
            65,058 & 
            18 \\
        \midrule
        EB-NeRD &
            15,143 &
            15,339 & 
            111.72 &
            3,732,517 &
            11,421 &
            23 \\
        \midrule
        NeMig &
            3,242 &
            3,242 & 
            5.78 &
            97,232 &
            4,933 &
            26 \\
        \bottomrule
    \end{tabular}
    }
    \Description{Comparison of EB-NeRD, MIND, and NeMig in terms of users, average number or articles in a user history, impressions, articles, and unique article categories.}
    \label{tab:datasets}
\end{table}

\paragraph{\textbf{Models:}}
The experiment includes LSTUR, NRMS, NPA,\footnote{We did not fine-tune the models, but used the hyperparameters from the official repository: \url{https://github.com/recommenders-team/recommenders}}
\rdw{}, RWE-D, and the norm-aware \drdw.\footnote{We perform $3$ hops, with the exception of \drdw{} on NeMig using $5$ hops, as the graph is too sparse to give us $20$ items with a lower number of hops.}
Furthermore, we used the filtering algorithms PLD and EPD.
A random selection of articles (RND) is used as a baseline for comparison.
We calculate the top $20$ item recommendations for each user, using reference values for list sizes from the literature~\cite{vrijenhoek2022radio,heitz2024recommendations}.
The word embeddings for the neural models are GloVe\footnote{GloVe word vectors: \url{https://nlp.stanford.edu/projects/glove/}} for MIND as well as fastText for both EB-NeRD (Danish) and NeMig (German).\footnote{fastText word vectors: \url{https://fasttext.cc/docs/en/crawl-vectors.html}}
We used off-the-shelf models to calculate article similarity for mapping cold items to users for random walks.
The sentence transformers in our workflow include 
RoBERTa\footnote{RoBERTa model: \url{https://huggingface.co/FacebookAI/roberta-base}} for EB-NeRD, MPNet\footnote{MPNet model: \url{https://huggingface.co/sentence-transformers/all-mpnet-base-v2}} for MIND, and E5\footnote{E5 model: \url{https://huggingface.co/intfloat/multilingual-e5-base}} for NeMig.
For \drdw, NTD covers the dimension of political parties and article sentiment.
By default, NTD consists of five buckets:
1) governing parties ($15\%$), 
2) opposition parties ($15\%$), 
3) governing \textit{and} opposition parties ($15\%$),
4) others (e.g., independent and foreign parties, $15\%$), and 
5) articles with no political party mentions ($40\%$).
To ensure a broad coverage, the value ranges and percentages for the sentiment distribution are $[-1,-0.5)$ ($20\%$), $[-0.5,0)$ ($30\%$), $[0,0.5)$ ($30\%$), $[0.5,1]$ ($20\%$) for EB-NeRD and MIND, and $[-1,0)$ ($50\%$), $[0,1]$ ($50\%$) for NeMig.\footnote{For NeMig, we used only positive and negative sentiments, as the data did not allow for a more detailed assessment of sentiment.}

\begin{table*}

    \caption{Overview for \textit{EB-NeRD} of the diversity scores for the top 20 news recommendations and AUC for item prediction scores of the entire article pool. Best values are highlighted in \colorbox{green!17}{green}, second best in \colorbox{blue!17}{blue}, and third best in \colorbox{orange!17}{orange}.}
    \scalebox{\width}{
       
        \begin{tabular}{lc|cccccc|cccccc|c}

            \toprule
            
            \textbf{Model} &
                \parbox{1.6cm}{\centering{\textbf{Re-ranking}}} &
                \parbox{1.0cm}{\centering{\textbf{Activ.}}} &
                \parbox{1.0cm}{\centering{\textbf{Cat. Calib.}}} &
                \parbox{1.0cm}{\centering{\textbf{Comp.  Calib.}}} &
                \parbox{1.0cm}{\centering{\textbf{Frag.}}} &
                \parbox{1.0cm}{\centering{\textbf{Alt. Voices}}} &
                \parbox{1.0cm}{\centering{\textbf{Repr.}}} &
                \parbox{1.0cm}{\centering{\textbf{Cat. Gini}}} &
                \parbox{1.0cm}{\centering{\textbf{Sent. Gini}}} &
                \parbox{1.0cm}{\centering{\textbf{Party Gini}}} &
                \parbox{1.0cm}{\centering{C\textbf{at. ILD}}} &
                \parbox{1.0cm}{\centering{\textbf{Sent. ILD}}} &
                \parbox{1.0cm}{\centering{\textbf{Party ILD}}} &
                \parbox{1.0cm}{\centering{\textbf{AUC}}} \\

            \midrule

            NTV &
                 &
                1.000 &
                0.000 &
                0.000 &
                0.000 &
                0.000 &
                1.000 &
                0.000 &
                0.133 &
                0.250 &
                1.000 &
                0.779 &
                0.789 &
                1.000 \\

            \midrule
        
            LSTUR~\cite{an2019lstur} &
                 &
                0.190 &
                \cellcolor{orange!17}0.433 &
                0.257 &
                0.610 &
                0.063 &
                0.372 &
                0.832 &
                0.645 &
                0.874 &
                0.740 &
                0.552 &
                0.343 &
                \cellcolor{blue!17}0.564 \\

            NPA~\cite{wu2019npa} &
                 &
                0.202 &
                0.468 &
                0.266 &
                0.619 &
                0.061 &
                0.365 &
                0.826 & 
                0.605 &
                0.877 &
                0.753 & 
                0.580 &
                0.337 &
                \cellcolor{orange!17}0.554 \\

            NRMS~\cite{wu2019nrms} &
                 &
                0.204 &
                0.509 &
                0.285 &
                0.632 &
                \cellcolor{orange!17}0.060 &
                0.362 &
                \cellcolor{orange!17}0.791 &
                0.544 &
                0.880 &
                0.794 &
                0.624 &
                0.326 &
                0.549 \\

            \midrule

            LSTUR &
                G-KL &
                0.282 &
                0.444 &
                0.240 &
                0.612 &
                0.104 &
                0.546 &
                0.828 &
                \cellcolor{green!17}0.133 &
                \cellcolor{green!17}0.250 &
                0.754 &
                \cellcolor{green!17}0.779 &
                \cellcolor{green!17}0.789 &
                 \\

            &
                PM-2 &
                0.284 &
                0.440 &
                0.244 &
                0.604 &
                0.115 &
                0.546 &
                0.819 &
                \cellcolor{blue!17}0.150 &
                \cellcolor{green!17}0.250 &
                0.766 &
                \cellcolor{blue!17}0.776 &
                \cellcolor{green!17}0.789 &
                 \\
                
            &
                MMR &
                0.295 &
                \cellcolor{orange!17}0.433 &
                0.235 &
                0.613 &
                0.118 &
                0.575 &
                0.823 &
                0.226 &
                \cellcolor{blue!17}0.270 &
                0.759 &
                0.762 &
                \cellcolor{blue!17}0.807 &
                 \\

            &
                POS &
                0.365 &
                0.480 &
                0.259 &
                0.667 &
                0.110 &
                \cellcolor{orange!17}0.731 &
                0.844 &
                0.759 &
                0.853 &
                0.699 &
                0.430 &
                0.344 &
                 \\

            &
               ATT  &
                \cellcolor{orange!17}0.371 &
                0.476 &
                0.258 &
                0.667 &
                0.113 &
                \cellcolor{green!17}0.743 &
                0.840 &
                0.754 &
                0.850 &
                0.712 &
                0.436 &
                0.350 &
                 \\

            \midrule

            NPA &
                G-KL &
                0.292 &
                0.469 &
                0.237 &
                0.589 &
                0.108 &
                0.545 &
                0.810 &
                \cellcolor{green!17}0.133 &
                \cellcolor{green!17}0.250 &
                0.775 &
                \cellcolor{green!17}0.779 &
                \cellcolor{green!17}0.789 &
                 \\

             &
                PM-2 &
                0.302 &
                0.471 &
                0.239 &
                0.576 &
                0.122 &
                0.547 &
                0.806 &
                \cellcolor{blue!17}0.150 &
                \cellcolor{green!17}0.250 &
                0.783 &
                \cellcolor{blue!17}0.776 &
                \cellcolor{green!17}0.789 &
                 \\
            
            &
                MMR &
                0.314 &
                0.463 &
                0.236 &
                0.598 &
                0.110 &
                0.575 &
                0.808 &
                0.219 &
               \cellcolor{blue!17} 0.270 &
                0.772 &
                0.763 &
                \cellcolor{blue!17}0.807 &
                 \\

            &
                POS &
                0.370 &
                0.487 &
                0.263 &
                0.656 &
                0.108 &
                0.727 &
                0.841 &
                0.755 &
                0.863 &
                0.707 &
                0.431 &
                0.327 &
                 \\

            &
                ATT &
                \cellcolor{green!17}0.377 &
                0.484 &
                0.259 &
                0.655 &
                0.111 &
                \cellcolor{blue!17}0.742 &
                0.841 &
                0.742 &
                0.860 &
                0.708 &
                0.448 &
                0.331 &
                 \\

            \midrule

            NRMS &
                G-KL &
                0.307 &
                0.482 &
                0.239 &
                0.608 &
                0.091 &
                0.544 &
                0.798 &
                \cellcolor{green!17}0.133 &
                \cellcolor{green!17}0.250 &
                0.790 &
                \cellcolor{green!17}0.779 &
                \cellcolor{green!17}0.789 &
                 \\

             &
                PM-2 &
                0.316 &
                0.483 &
                0.244 &
                0.598 &
                0.099 &
                0.544 &
                0.796 &
                \cellcolor{blue!17}0.150 &
                \cellcolor{green!17}0.250 &
                0.794 &
                \cellcolor{blue!17}0.776 &
                \cellcolor{green!17}0.789 &
                 \\

            &
                MMR &
                0.315 &
                0.477 &
                0.238 &
                0.616 &
                0.099 &
                0.571 &
                0.798 &
                \cellcolor{orange!17}0.204 &
                \cellcolor{blue!17}0.270 &
                0.790 &
                \cellcolor{orange!17}0.767 &
                \cellcolor{blue!17}0.807 &
                 \\

            &
                POS &
                0.366 &
                0.480 &
                0.263 &
                0.657 &
                0.110 &
                \cellcolor{orange!17}0.731 &
                0.841 &
                0.765 &
                0.851 &
                0.701 &
                0.419 &
                0.344 &
                 \\

            &
                ATT &
                \cellcolor{orange!17}0.371 &
                0.476 &
                0.260 &
                0.655 &
                0.112 &
                \cellcolor{blue!17}0.742 &
                0.839 &
                0.747 &
                0.848 &
                0.709 &
                0.444 &
                0.349 &
                 \\

            \midrule

            \drdw{} &
                 &
                \cellcolor{blue!17}0.374 &
                \cellcolor{green!17}0.407 &
                \cellcolor{orange!17}0.229 &
                \cellcolor{green!17}0.394 &
                0.107 &
                0.556 &
                0.798 &
                \cellcolor{green!17}0.133 &
                \cellcolor{green!17}0.250 &
                \cellcolor{blue!17}0.810 &
                \cellcolor{green!17}0.779 &
                \cellcolor{green!17}0.789 &
                \cellcolor{orange!17}0.554 \\

            \rdw{}~\cite{christoffel2015blockbusters} &
                 &
                0.222 &
                \cellcolor{blue!17}0.415 &
                0.235 &
                0.582 &
                0.080 &
                0.376 &
                0.840 &
                0.755 &
                0.856 &
                0.743 &
                0.439 &
                0.392 &
                \cellcolor{green!17}0.565 \\

            RWE-D~\cite{paudel2021random} &
                 &
                0.256 &
                0.435 &
                \cellcolor{blue!17}0.222 &
                \cellcolor{orange!17}0.443 &
                0.100 &
                0.372 &
                0.857 &
                0.802 & 
                0.842 &
                0.735 &
                0.377 & 
                0.433 &
                \cellcolor{orange!17}0.554 \\

            \midrule

            PLD~\cite{heitz2022benefits} &
                 &
                0.152 &
                0.459 &
                0.268 &
                \cellcolor{blue!17}0.418 &
                \cellcolor{green!17}0.038 &
                0.432 &
                0.801 &
                0.687 &
                0.749 &
                0.782 &
                0.534 &
                0.556 &
                 \\

            EPD~\cite{heitz2023deliberative} &
                 &
                0.139 &
                0.443 &
                \cellcolor{green!17}0.207 &
                0.486 &
                0.081 &
                0.505 &
                \cellcolor{blue!17}0.773 &
                0.611 &
                \cellcolor{orange!17}0.667 &
                \cellcolor{orange!17}0.802 &
                0.577 &
                \cellcolor{orange!17}0.610 &
                 \\
            
            \midrule
            
            Random &
                 & 
                0.180 &
                0.461 &
                0.256 &
                0.705 &
                \cellcolor{blue!17}0.054 &
                0.366 &
                \cellcolor{green!17}0.756 &
                0.634 &
                0.873 &
                \cellcolor{green!17}0.842 &
                0.564 &
                0.346 &
                0.500 \\

            \bottomrule
            
        \end{tabular}

    }
    
    \Description{Overview for EB-NeRD of the diversity scores for the top 20 news recommendations and AUC scores for predicting user impressions.}
    \label{tab:results_ebnerd}
    
\end{table*}

\begin{table*}

    \caption{Overview for \textit{MIND} of the diversity scores for the top 20 news recommendations and AUC for item prediction scores of the entire article pool. Best values are highlighted in \colorbox{green!17}{green}, second best in \colorbox{blue!17}{blue}, and third best in \colorbox{orange!17}{orange}.}
    \scalebox{\width}{
       
        \begin{tabular}{lc|cccccc|cccccc|c}

            \toprule
            
            \textbf{Model} &
                \parbox{1.6cm}{\centering{\textbf{Re-ranking}}} &
                \parbox{1.0cm}{\centering{\textbf{Activ.}}} &
                \parbox{1.0cm}{\centering{\textbf{Cat. Calib.}}} &
                \parbox{1.0cm}{\centering{\textbf{Comp.  Calib.}}} &
                \parbox{1.0cm}{\centering{\textbf{Frag.}}} &
                \parbox{1.0cm}{\centering{\textbf{Alt. Voices}}} &
                \parbox{1.0cm}{\centering{\textbf{Repr.}}} &
                \parbox{1.0cm}{\centering{\textbf{Cat. Gini}}} &
                \parbox{1.0cm}{\centering{\textbf{Sent. Gini}}} &
                \parbox{1.0cm}{\centering{\textbf{Party Gini}}} &
                \parbox{1.0cm}{\centering{C\textbf{at. ILD}}} &
                \parbox{1.0cm}{\centering{\textbf{Sent. ILD}}} &
                \parbox{1.0cm}{\centering{\textbf{Party ILD}}} &
                \parbox{1.0cm}{\centering{\textbf{AUC}}} \\

            \midrule

            NTV &
                 &
                1.000 &
                0.000 &
                0.000 &
                0.000 &
                0.000 &
                1.000 &
                0.000 &
                0.133 &
                0.250 &
                1.000 &
                0.779 &
                0.789 &
                1.000 \\
                
            \midrule
        
            LSTUR~\cite{an2019lstur} &
                 &
                0.266 &
                0.620 &
                0.313 &
                0.503 &
                \cellcolor{green!17}0.051 &
                0.296 &
                0.855 &
                0.593 &
                0.905 &
                0.618 &
                0.585 &
                0.247 &
                \cellcolor{orange!17}0.593 \\

            NPA~\cite{wu2019npa} &
                 &
                0.191 &
                0.574 &
                0.329 &
                \cellcolor{orange!17}0.443 &
                0.076 &
                0.280 &
                0.792 & 
                0.614 &
                0.854 &
                0.751 & 
                0.587 &
                0.323 &
                \cellcolor{blue!17}0.595 \\

            NRMS~\cite{wu2019nrms} &
                 &
                0.211 &
                \cellcolor{blue!17}0.559 &
                0.313 &
                0.712 &
                0.071 &
                0.305 &
                0.756 &
                0.554 &
                0.896 &
                0.764 &
                0.615 &
                0.246 &
                \cellcolor{green!17}0.626 \\

            \midrule

            LSTUR &
                G-KL &
                0.250 &
                0.619 &
                \cellcolor{orange!17}0.311 &
                0.558 &
                0.097 &
                0.408 &
                0.786 &
                \cellcolor{green!17}0.133 &
                \cellcolor{green!17}0.250 &
                0.769 &
                \cellcolor{green!17}0.779 &
                \cellcolor{green!17}0.789&
                 \\

            &
                PM-2 &
                0.227 &
                0.607 &
                0.317 &
                0.574 &
                0.086 &
                0.418 &
                0.809 &
                \cellcolor{green!17}0.133 &
                \cellcolor{green!17}0.250 &
                0.734 &
                \cellcolor{green!17}0.779 &
                \cellcolor{green!17}0.789&
                 \\
                
            &
                MMR &
                0.287 &
                0.604 &
                0.334 &
                0.594 &
                0.102 &
                0.471 &
                0.766 &
                \cellcolor{blue!17}0.000 &
                \cellcolor{blue!17}0.000 &
                0.781 &
                \cellcolor{blue!17}0.789 &
                \cellcolor{blue!17}0.842 &
                 \\

            &
                POS &
                0.299 &
                0.631 &
                0.338 &
                0.693 &
                0.120 &
                0.538 &
                0.790 &
                0.777 &
                0.623 &
                0.693 &
                0.393 &
                0.642 &
                 \\

            &
                ATT &
                0.289 &
                0.633 &
                0.340 &
                0.688 &
                0.134 &
                0.544 &
                0.786 &
                0.746 &
                0.618 &
                0.712 &
                0.438 &
                \cellcolor{orange!17}0.652 &
                 \\

            \midrule

             NPA &
                G-KL &
                0.226 &
                0.585 &
                0.327 &
                0.458 &
                0.098 &
                0.434 &
                0.829 & 
                \cellcolor{green!17}0.133 &
                \cellcolor{green!17}0.250 &
                0.691 &
                \cellcolor{green!17}0.779 &
                \cellcolor{green!17}0.789&
                 \\

             &
                PM-2 &
                0.251&
                0.599 &
                0.331 &
                0.465 &
                0.117 &
                0.438 &
                0.858 &
                \cellcolor{green!17}0.133 &
                \cellcolor{green!17}0.250 &
                0.651 &
                \cellcolor{green!17}0.779 &
                \cellcolor{green!17}0.789 &
                 \\
            
            &
                MMR &
                0.320 &
                0.581 &
                0.329 &
                0.475 &
                0.137 &
                0.517 &
                0.837 &
                \cellcolor{blue!17}0.000 &
                \cellcolor{blue!17}0.000 &
                0.691 &
                \cellcolor{blue!17}0.789 &
                \cellcolor{blue!17}0.842 &
                 \\

            &
                POS &
                0.329 &
                0.636 &
                0.372 &
                0.549 &
                0.173 &
                \cellcolor{green!17}0.568 &
                0.900 &
                0.745 &
                0.640 &
                0.436 &
                0.431 &
                0.637 &
                 \\

            &
                ATT &
                \cellcolor{orange!17}0.331 &
                0.639 &
                0.372 &
                0.556 &
                0.171 &
                \cellcolor{orange!17}0.563 &
                0.896 &
                0.740 &
                0.649 &
                0.449 &
                0.439 &
                0.628 &
                 \\

           \midrule
            
           NRMS &
                G-KL &
                0.232 &
                0.563 &
                0.315 &
                0.692 &
                0.075 &
                0.402 &
                \cellcolor{orange!17}0.713 &
                \cellcolor{green!17}0.133 &
                \cellcolor{green!17}0.250 &
                0.812 &
                \cellcolor{green!17}0.779 &
                \cellcolor{green!17}0.789 &
                 \\

             &
                PM-2 &
                0.238 &
                \cellcolor{orange!17}0.560 &
                \cellcolor{blue!17}0.310 &
                0.706 &
                0.079 &
                0.390 &
                0.729 &
                \cellcolor{green!17}0.133 &
                \cellcolor{green!17}0.250 &
                0.801 &
                \cellcolor{green!17}0.779 &
                \cellcolor{green!17}0.789&
                 \\

            &
                MMR &
                0.284 &
                0.574 &
                0.322 &
                0.692 &
                0.088 &
                0.468 &
                0.716 &
                \cellcolor{blue!17}0.000 &
                \cellcolor{blue!17}0.000 &
                0.809 &
                \cellcolor{blue!17}0.789 &
                \cellcolor{blue!17}0.842 &
                 \\

            &
                POS &
                \cellcolor{green!17}0.346 &
                0.605 &
                0.352 &
                0.713 &
                0.117 &
                0.561 &
                0.766 &
                0.729 &
                0.623 &
                0.740 &
                0.435 &
                0.639 &
                 \\

            &
                ATT &
                \cellcolor{blue!17}0.333 &
                0.603 &
                0.352 &
                0.712 &
                0.119 &
                \cellcolor{blue!17}0.567 &
                0.770 &
                0.713 &
                \cellcolor{orange!17}0.613 &
                0.736 &
                0.461 &
                0.649 &
                 \\

            \midrule

     
            \drdw{} &
                 &
                0.279 &
                0.592 &
                \cellcolor{blue!17}0.310 &
                \cellcolor{blue!17}0.433 &
                0.097 &
                0.419 &
                0.766 &
                \cellcolor{green!17}0.133 &
                \cellcolor{green!17}0.250 &
                0.761 &
                \cellcolor{green!17}0.779 &
                \cellcolor{green!17}0.789 &
                0.511 \\

            \rdw{}~\cite{christoffel2015blockbusters} &
                 &
                0.215 &
                \cellcolor{green!17}0.543 &
                0.312 &
                0.709 &
                0.074 &
                0.308 &
                0.737 &
                0.540 &
                0.904 &
                0.783 &
                0.622 &
                0.230 &
                0.532 \\

            RWE-D~\cite{paudel2021random} &
                 &
                0.225 &
                0.583 &
                \cellcolor{green!17}0.304 &
                \cellcolor{green!17}0.398 &
                0.102 &
                0.298 &
                0.724 &
                \cellcolor{orange!17}0.364 &
                0.830 &
                0.805 &
                \cellcolor{orange!17}0.715 &
                0.352 &
                0.512 \\

            \midrule

            PLD~\cite{heitz2022benefits} &
                 &
                0.146 &
                0.580 &
                0.345 &
                0.560 &
                \cellcolor{orange!17}0.059 &
                0.336 &
                \cellcolor{green!17}0.658 &
                0.484 &
                0.795 &
                \cellcolor{blue!17}0.857 &
                0.665 & 
                0.403 &
                 \\

            EPD~\cite{heitz2023deliberative} &
                 &
                0.274 & 
                0.614 &
                0.318 &
                0.446 &
                0.100 &
                0.399 &
                0.726 &
                0.533 &
                0.850 &
                \cellcolor{orange!17}0.823 &
                0.626 & 
                0.377 &
                 \\
            
            \midrule

            Random &
                 & 
                0.197 &
                0.618 &
                0.314 &
                0.702 &
                \cellcolor{blue!17}0.057 &
                0.301 &
                \cellcolor{blue!17}0.663 &
                0.503 &
                0.897 &
                \cellcolor{green!17}0.861 &
                0.649 &
                0.251 &
                0.498 \\

            \bottomrule
            
        \end{tabular}

    }
    
    \Description{Overview for MIND of the diversity scores for the top 20 news recommendations and AUC scores for predicting user impressions.}
    \label{tab:results_mind}
    
\end{table*}

\begin{table*}

    \caption{Overview for \textit{NeMig} of the diversity scores for the top 20 news recommendations and AUC for item prediction scores of the entire article pool. Best values are highlighted in \colorbox{green!17}{green}, second best in \colorbox{blue!17}{blue}, and third best in \colorbox{orange!17}{orange}.}
    \scalebox{\width}{
       
        \begin{tabular}{lc|cccccc|cccccc|c}

            \toprule
            
            \textbf{Model} &
                \parbox{1.6cm}{\centering{\textbf{Re-ranking}}} &
                \parbox{1.0cm}{\centering{\textbf{Activ.}}} &
                \parbox{1.0cm}{\centering{\textbf{Cat. Calib.}}} &
                \parbox{1.0cm}{\centering{\textbf{Comp.  Calib.}}} &
                \parbox{1.0cm}{\centering{\textbf{Frag.}}} &
                \parbox{1.0cm}{\centering{\textbf{Alt. Voices}}} &
                \parbox{1.0cm}{\centering{\textbf{Repr.}}} &
                \parbox{1.0cm}{\centering{\textbf{Cat. Gini}}} &
                \parbox{1.0cm}{\centering{\textbf{Sent. Gini}}} &
                \parbox{1.0cm}{\centering{\textbf{Party Gini}}} &
                \parbox{1.0cm}{\centering{C\textbf{at. ILD}}} &
                \parbox{1.0cm}{\centering{\textbf{Sent. ILD}}} &
                \parbox{1.0cm}{\centering{\textbf{Party ILD}}} &
                \parbox{1.0cm}{\centering{\textbf{AUC}}} \\

            \midrule

            NTV &
                 &
                1.000 &
                0.000 &
                0.000 &
                0.000 &
                0.000 &
                1.000 &
                0.000 &
                0.000 &
                0.250 &
                1.000 &
                0.526 &
                0.789 &
                1.000 \\
            
            \midrule
        
            LSTUR~\cite{an2019lstur} &
                 &
                0.192 &
                0.637 &
                0.415 &
                0.870 &
                0.044 &
                0.545 &
                \cellcolor{blue!17}0.850 &
                0.951 &
                0.752 &
                \cellcolor{blue!17}0.740 &
                0.048 &
                0.517 &
                \cellcolor{orange!17}0.535 \\

            NPA~\cite{wu2019npa} &
                 &
                0.206 &
                0.601 &
                0.420 &
                0.787 &
                0.073 &
                0.568 &
                0.905 & 
                0.980 &
                0.725 &
                0.577 & 
                0.019 &
                0.556 &
                0.491 \\

            NRMS~\cite{wu2019nrms} &
                 &
                0.220 &
                0.629 &
                0.428 &
                0.730 &
                0.055 &
                0.556 &
                0.864 &
                0.963 &
                0.808 &
                0.707 &
                0.037 &
                0.468 &
                \cellcolor{green!17}0.552 \\

            \midrule

            LSTUR &
                G-KL &
                0.282 &
                0.604 &
                0.385 &
                0.758 &
                0.045 &
                0.565 &
                0.904 &
                \cellcolor{green!17}0.000 &
                \cellcolor{green!17}0.250 &
                0.618 &
                \cellcolor{green!17}0.526 &
                \cellcolor{green!17}0.789 &
                 \\

            &
                PM-2 &
                0.288 &
                0.605 &
                0.385 &
                0.763 &
                0.046 &
                0.565 &
                0.902 &
                \cellcolor{green!17}0.000 &
                \cellcolor{green!17}0.250 &
                0.622 &
                \cellcolor{green!17}0.526 &
                \cellcolor{green!17}0.789 &
                 \\
                
            &
                MMR &
                0.192 &
                0.617 &
                0.400 &
                0.826 &
                0.046 &
                0.535 &
                0.876 &
                \cellcolor{blue!17}0.497 &
                \cellcolor{orange!17}0.431 &
                0.675 &
                \cellcolor{blue!17}0.396 &
                \cellcolor{orange!17}0.723 &
                 \\

            &
                POS &
                0.535 &
                0.602 &
                0.371 &
                0.651 &
                0.076 &
                0.614 &
                0.898 &
                0.927 &
                0.667 &
                0.619 &
                0.048 &
                0.552 &
                 \\

            &
                ATT &
                \cellcolor{blue!17}0.543 &
                0.605 &
                \cellcolor{blue!17}0.364 &
                \cellcolor{blue!17}0.576 &
                0.078 &
                0.611 &
                0.896 &
                0.967 &
                0.652 &
                0.632 &
                0.022 &
                0.579 &
                 \\

            \midrule

             NPA &
                G-KL &
                0.275 &
                \cellcolor{blue!17}0.598 &
                0.385 &
                0.686 &
                0.042 &
                0.571 &
                0.910 &
                \cellcolor{green!17}0.000 &
                \cellcolor{green!17}0.250 &
                0.572 &
                \cellcolor{green!17}0.526 &
                \cellcolor{green!17}0.789 &
                 \\

             &
                PM-2 &
                0.285 &
                0.600 &
                0.386 &
                0.693 &
                0.045 &
                0.571 &
                0.906 &
                \cellcolor{green!17}0.000 &
                \cellcolor{green!17}0.250 &
                0.582 &
                \cellcolor{green!17}0.526 &
                \cellcolor{green!17}0.789 &
                 \\
            
            &
                MMR &
                0.234 &
                \cellcolor{green!17}0.594 &
                0.398 &
                0.736 &
                0.056 &
                0.549 &
                0.913 &
                \cellcolor{orange!17}0.499 &
                \cellcolor{blue!17}0.415 &
                0.548 &
                \cellcolor{orange!17}0.395 &
                \cellcolor{blue!17}0.739 &
                 \\

            &
                POS &
                \cellcolor{orange!17}0.540 &
                \cellcolor{orange!17}0.599 &
                0.373 &
                0.664 &
                0.080 &
                0.615 &
                0.902 &
                0.923 &
                0.707 &
                0.608 &
                0.051 &
                0.510 &
                 \\

            &
                ATT &
                0.538 &
                0.603 &
                0.368 &
                0.588 &
                0.090 &
                \cellcolor{orange!17}0.616 &
                0.899 &
                0.967 &
                0.701 &
                0.628 &
                0.022 &
                0.534 &
                 \\

           \midrule
            
           NRMS &
                G-KL &
                0.281 &
                0.624 &
                0.382 &
                0.667 &
                \cellcolor{orange!17}0.040 &
                0.576 &
                0.877 &
                \cellcolor{green!17}0.000 &
                \cellcolor{green!17}0.250 &
                0.680 &
                \cellcolor{green!17}0.526 &
                \cellcolor{green!17}0.789 &
                 \\

             &
                PM-2 &
                0.289 &
                0.622 &
                0.382 &
                0.664 &
                \cellcolor{blue!17}0.037 &
                0.575 &
                0.879 &
                \cellcolor{green!17}0.000 &
                \cellcolor{green!17}0.250 &
                0.675 &
                \cellcolor{green!17}0.526 &
                \cellcolor{green!17}0.789 &
                 \\

            &
                MMR &
                0.219 &
                0.626 &
                0.411 &
                0.703 &
                0.070 &
                0.552 &
                0.870 &
                0.500 &
                0.452 &
                0.685 &
                \cellcolor{orange!17}0.395 &
                0.705 &
                 \\

            &
                POS &
                \cellcolor{orange!17}0.540 &
                0.611 &
                \cellcolor{orange!17}0.367 &
                0.609 &
                0.086 &
                \cellcolor{blue!17}0.620 &
                0.877 &
                0.935 &
                0.716 &
                0.671 &
                0.043 &
                0.510 &
                 \\

            &
                ATT &
                \cellcolor{green!17}0.546 &
                0.618 &
                \cellcolor{green!17}0.363 &
                \cellcolor{green!17}0.521 &
                0.097 &
                \cellcolor{green!17}0.625 &
                0.870 &
                0.979 &
                0.739 &
                0.693 &
                0.014 &
                0.500 &
                 \\

            \midrule

            \drdw{} &
                 &
                0.289 &
                \cellcolor{blue!17}0.598 &
                0.382 &
                0.737 &
                0.042 &
                0.554 &
                0.887 &
                \cellcolor{green!17}0.000 &
                \cellcolor{green!17}0.250 &
                0.653 &
                \cellcolor{green!17}0.526 &
                \cellcolor{green!17}0.789 &
                \cellcolor{blue!17}0.550 \\

            \rdw{}~\cite{christoffel2015blockbusters} &
                 &
                0.185 &
                0.612 &
                0.408 &
                0.880 &
                0.049 &
                0.545 &
                \cellcolor{orange!17}0.856 &
                0.921 &
                0.761 &
                0.717 &
                0.076 &
                0.508 &
                0.448 \\

            RWE-D~\cite{paudel2021random} &
                 &
                0.186 &
                0.633 &
                0.416 &
                0.877 &
                0.055 &
                0.554 &
                \cellcolor{blue!17}0.850 &
                0.943 &
                0.801 &
                \cellcolor{green!17}0.762 &
                0.055 &
                0.445 &
                0.451 \\

            \midrule

            PLD~\cite{heitz2022benefits} &
                 &
                0.127 &
                0.625 &
                0.417 &
                0.641 &
                \cellcolor{green!17}0.027 &
                0.526 &
                0.858 &
                0.926 &
                0.792 &
                0.716 &
                0.074 &
                0.466 &
                 \\

            EPD~\cite{heitz2023deliberative} &
                 &
                0.218 &
                0.615 &
                0.433 &
                \cellcolor{orange!17}0.579 &
                0.058 &
                0.587 &
                0.876 &
                0.900 &
                0.733 &
                0.660 &
                0.097 &
                0.558 &
                 \\
            
            \midrule

            Random &
                 & 
                0.185 &
                0.628 &
                0.412 &
                0.879 &
                0.047 &
                0.541 &
                \cellcolor{green!17}0.843 &
                0.929 &
                0.772 &
                \cellcolor{orange!17}0.738 &
                0.068 &
                0.489 &
                0.498 \\

            \bottomrule
            
        \end{tabular}

    }
    
    \Description{Overview for NeMig of the diversity scores for the top 20 news recommendations and AUC scores for predicting user impressions.}
    \label{tab:results_nemig}
    
\end{table*}

\paragraph{\textbf{Re-rankers:}}
We experiment with the three static re-rankers mentioned in Section~\ref{sec:stage-3}, namely Greedy-KL (G-KL), PM-2, and MMR, to calculate the top $20$ recommendations for each user.
G-KL and PM-2 re-rankers use the same diversity dimensions and distributions as those defined by our aforementioned target distribution.
More precisely, the diversity dimensions include sentiment and political parties, with equal weights given to both dimensions.

We added the DAP dynamic re-ranking strategy that simulated a user with strong position preferences (clicking predominantly on the top-most items, POS) and one reflecting attribute preferences for political parties and sentiment (ATT).

\paragraph{\textbf{Metrics:}}
We use five diversity metrics for measuring divergence of Activation (Activ.), Category Calibration (Cat. Calib.), Complexity Calibration (Comp. Calib.), Fragmentation (Frag.), Alternative Voices (Alt. Voices), and Representation (Repr.) adopted from the RADio metrics~\cite{vrijenhoek2022radio}.\footnote{We applied Jensen-Shannon (JS) divergence without incorporating rank-awareness.}
We use the test set impression item pool as a reference distribution for calculating the Activation, Alternative Voices, and Representation metric.
Calibration compares a user’s recommendation distribution with their own reading history, and Fragmentation compares the history with a randomly sampled user.

We use two traditional diversity metrics, namely the Gini coefficient for article category (Cat. Gini), sentiment (Sent. Gini), and political parties (Party Gini), together with intra-list distance for article category (Cat. ILD), sentiment (Sent. ILD), and political parties (Party ILD).
ILD was computed using one-hot encoded vectors representing sentiment categories and political party mentions.

AUC is used for the accuracy assessment.\footnote{Our focus here is to demonstrate the newly added diversity metrics. Further accuracy metrics are omitted from the analysis, as they are already part of the Cornac codebase.} 
The computation is based on pairwise comparisons between predicted scores for classifying clicked and unclicked items in the user impressions.
Given the generally low values for AUC, we chose not to recompute the AUC for the re-rankers.
We acknowledge, however, that one can also compute AUC on ranked lists and plan to make Informfully Recommenders configurable in that respect in the future. 



\section{Results and Discussion}
\label{sec:results_and_discussion}
Table~\ref{tab:results_ebnerd} shows the values for the EB-NeRD dataset, Table~\ref{tab:results_mind} for MIND, and Table~\ref{tab:results_nemig} for NeMig.
Splitting the RS pipeline into separate stages allows us to look at the impact of each element: the dataset, the recommender models, and the applied re-ranking techniques on target metrics.
We compare the families of approaches:
1) traditional neural models (LSTUR, NPA, NRMS), 
2) baseline random walk models (\rdw and RWE-D), and
3) NTD-optimizing algorithms (\drdw, PLD, and EPD) and re-rankers.
A separate row for Normative Target Values (NTV) is included that shows the best achievable scores given the underlying dataset and NTD.

\paragraph{\textbf{RADio Diversity Metrics:}}
RADio metrics assess RS performance on the basis of a divergence between an individual's recommendations and the overall item pool (cf.~\cite{vrijenhoek2022radio}). 
%
For \textbf{EB-NeRD}, we see that target distribution-optimizing models achieve the top score in all but one category.
The performance increase over neural models and re-ranking is especially large for fragmentation and alternative voices. 
\drdw{} creates the least fragmented readership, and PLD gives the most exposure to minority positions.
%
A similar picture presents itself with \textbf{MIND}.
But now we see that NPA and NRMS with dynamic position- and topic-aware re-ranking can score highest in terms of activation and representation.
The comparatively larger item pool of MIND (it contains more items than users, see Table~\ref{tab:datasets}) provides the dynamic approach with sufficient items to diversify the recommendations.
%
This holds true for \textbf{NeMig} as well.
Dynamic re-ranking for LSTUR, NPA, and NRMS outperforms most other approaches on RADio.
With NeMig being more than one order of magnitude smaller than EB-NeRD and MIND in terms of impressions, there does not seem to be enough data.
This also impacts random walk models, as a low number of impressions means a sparsely connected graph, explaining their poor performance.
%

\paragraph{\textbf{Traditional Diversity Metrics:}}
The Gini coefficient assesses equality.
The smaller the value, the more equal the distribution of a given attribute. 
ILD measures the average pairwise similarity between items using cosine distance.
The smaller the distance, the more similar the items within a list. 
Both Gini and ILD are frequently used as proxies for diversity~\cite{kunaver2017diversity}.
%
We see that the distribution-optimizing approaches with NTD (G-KL re-ranking and \drdw{} in particular) consistently reach the NTV for sentiment and party Gini as well as ILD across \textbf{all datasets}.
The same does not hold true for category Gini/ILD.
Part of the reason is that the underlying NTD of \drdw{} and parts of the re-rankers do not include the category.
To use, this presents evidence for the effectiveness of NTD-based approaches to tweak and adjust the article supply in the recommendation list.
%

\paragraph{\textbf{Accuracy and Diversity Tradeoffs}}
We calculate AUC to show that the target distribution-optimizing models and re-rankers not only diversify the recommendation list but also present relevant items similar to the state-of-the-art neural models.
For \textbf{EB-NeRD}, we see a tie in AUC for neural models and random walks.
Neural models substantially outperform the other two families on \textbf{MIND}. 
And the top spots for \textbf{NeMig} are shared between NRMS, \drdw, and LSTUR.
With the newly added metrics, our framework allows users to easily measure and study this \textit{accuracy-diversity tradeoff} (cf.~\cite{jannach2013recommenders,adomavicius2011improving}) between models.
Furthermore, \framework{} allows for assessing the \textit{diversity-diversity tradeoff} between traditional and normative versions.
This offers a useful alternative for cases where accuracy-focused metrics are not an ideal option for benchmarking and performance assessment (e.g., diversity models in news domain (cf.~\cite{heitz2024recommendations}).



\section{Limitations and Future Work}
\label{sec:limitations_and_future_work}
Our experiments are limited to offline news recommender systems benchmarking.
We chose to present experimental results in the news domain as this allowed us to exemplify the target distribution-optimizing capabilities of our diversity-aware framework extension.
Future work could focus on experimenting with different target distributions in other domains. 
In addition, we need to run online user studies to properly assess the impact of visualization (i.e., article position through varying item placement and accessibility through varying text complexity) on item consumption and engagement of diverse recommendations.



\section{Conclusion}
\label{sec:conclusion}
%
We present \framework{}, a reproducibility framework for recommender algorithms that facilitates diversity-driven offline benchmarking as well as online user experiments.
It provides a customizable end-to-end pipeline that allows for seamless algorithm development, benchmarking, and deployment.
Targeting user experiments, the pipeline includes text augmentation functionality, lightweight recommendation models, static and dynamic re-rankers, user simulators, normative and traditional diversity metrics, and item visualization.
\framework{} is a modular, diversity-driven extension to the well-established Cornac reproducibility framework.
%
We hope this framework enables researchers to incorporate social norms and values into their algorithms for production-ready recommenders to conduct user studies, as they present the ultimate test for any recommender system~\cite{jannach2020escaping}.
While we illustrated \framework{} in the news domain, it supports the development, benchmarking, \textit{and} deployment of end-to-end pipelines across different RS domains.

\balance

\begin{acks}
This work was partially funded by the Digital Society Initiative (DSI) of the University of Zurich (UZH) under a grant from the DSI Excellence Program, the Hasler Foundation, and the Swiss Federal Office of Communications (OFCOM).
We would like to thank our students Songyi Han, Shiyu Ran, Guanyu Chen, Qingyue Chen, Yujue Chen, and Lanlan Yang, who worked on this codebase as part of their coursework.
\end{acks}

\bibliographystyle{ACM-Reference-Format}
\bibliography{main}

\end{document}